\documentclass[%
 reprint,
 amsmath,amssymb,
 aps,
prb,
]{revtex4-2}
\DeclareUnicodeCharacter{2009}{\,} 
\usepackage{graphicx}
\usepackage{dcolumn}
\usepackage{bm}

\begin{document}
\title{Hydrodynamics in Semidilute Polyelectrolyte Solutions and Complex Coacervates}
\author{Shensheng Chen}
\author{Zhen-Gang Wang}%
\email{zgw@caltech.edu}
\affiliation{Division of Chemistry and Chemical Engineering, California Institute of Technology, Pasadena, CA 91125}%

\begin{abstract}
It is generally assumed that hydrodynamics in dense polyelectrolyte (PE) solutions, such as semidilute PE solutions and PE complex coacervates, is heavily screened and inconsequential. Here, using mesoscale molecular dynamics that explicitly accounts for hydrodynamics, we show that segmental dynamics in the subdiffusive regime show strong signatures of hydrodynamic interactions that persist well beyond the correlation length of semidilute PE solutions with moderately short chains. The strong hydrodynamic effects are also observed in coacervate systems containing moderately short chains, even with PE concentration as high as $30\%$. Our work fills a gap in the existing simulation literature on dense PE solutions and hints at the importance of hydrodynamics in the transport and rheological properties in broader polymer/polyelectrolyte solution systems.   
\end{abstract}

\maketitle

The assembly and dynamics of charged macromolecules in aqueous environments are ubiquitous aspects of soft matter that significantly impact  biological functions and various applications. For example, the layer-by-layer techniques that have revolutionized surface coating applications are based on the use of semidilute polyelectrolyte (PE) solutions \cite{Decher1992BuildupSurfaces}. Complexation of oppositely charged PEs and/or proteins in water can result in the formation of a dense liquid-like coacervate \cite{Alberti2019ConsiderationsCondensates},  which has attracted considerable attention in recent years due to its relevance to membraneless organelles in biology \cite{Hyman2014Liquid-LiquidBiology}, wet adhesion \cite{Dompe2019ThermoresponsiveAdhesive}, and biomedicine \cite{Blocher2017ComplexBiomedicine}. Understanding the dynamics of these dense PE solutions remains a difficult challenge in the study of soft matter physics, due to the complex interplay between electrostatics, polymer structure, and hydrodynamics \cite{Muthukumar201750thSolutions}.  

For PE solutions well above the overlap concentration, hydrodynamics is often considered to be heavily screened \cite{Rubinstein1994DynamicsSolutions,Dobrynin1995ScalingSolutions,Bollinger2021OverlapSolutions}, despite the fact that there is significant amount of solvent (water) ($> 90\%$ for typical semidilute PE solutions and $60\% \sim 90\%$ for PE complex coacervates). Because of this consideration and the high computational cost/complexity to include hydrodynamics, nearly all previous simulation studies \cite{Stevens1995TheStudy,Liao2007RouseStudy,Carrillo2011PolyelectrolytesSimulations,Bollinger2021OverlapSolutions,Carrillo2023Coarse-grainedSolutions,Yu2020CrossoverCoacervates,Liang2022AProperties} of semidilute PE solutions and complex coacervates neglect hydrodynamic interactions, by either treating the solvent implicitly, or treating solvent merely as packing particles. However, a recent work comparing experiment with Brownian dynamics simulation (without hydrodynamics) highlights the need to include the effects of hydrodynamic interactions in semidilute PE solutions to properly capture their dynamic behavior   \cite{Kumar2024PivotalSolutions}.

Without hydrodynamics, the dynamics of the polymers is expected to follow Rouse dynamics \cite{Rouse1953APolymers}, which is manifested in a monomer mean-squared distance (MSD) scaling with time as $\sim t^{1/2}$ in the subdiffusive regime. When the dynamics is dominated by hydrodynamics, which is usually considered to be the case in dilute solutions, the monomer MSD in the subdiffusive regime should follow Zimm dynamics \cite{Zimm1956DynamicsLoss} as $\sim t^{2/3}$. This difference also impacts the collective dynamics of the PE solutions, where the relaxation time extracted from the intermediate scattering function scales with wavenumber $q$ as $q^{-2}$ in Rouse dynamics and as $q^{-3}$ in Zimm dynamics. Given the important consequences of polymer dynamics on the transport and rheological response of semidilute PE solutions and coacervates, a systematic examination of the hydrodynamics effects in these dense PE solutions is warranted. 

In this work, we present results of computer simulation to show the importance of hydrodynamics in semidilute PE solutions and in PE complex coacervates under typical conditions.  The simulation employed in this work is based on our recently developed Gaussian-Core model with smeared electrostatics (GCMe)\cite{Ye2024GCMe:Systems}. Here we extend the GCMe model by including hydrodynamics. 

GCMe is a coarse-grained molecular dynamics simulation method where the excluded volume and electrostatic interactions are based on Gaussian mass and charge smearing \cite{Ye2024GCMe:Systems}, respectively. Briefly, in GCMe, coarse-grained beads (polymer, ion, water) $i$ and $j$ interact with a Gaussian soft-core repulsion of the form $u_{ex}(r_{ij})=A_{ij}\left (\frac{3}{2\pi\sigma_{ij}^2} \right)^{3/2} \exp(-\frac{3}{2\sigma_{ij}^2}r_{ij}^2)$, where $r_{ij}$ is the inter-particle separation distance, $A_{ij}$ is the strength of the repulsion, and $\sigma_{ij}=\sqrt{\sigma_i^2+\sigma_j^2}$ with $\sigma_i$ and $\sigma_j$ being the mass smear radius of particles $i$ and $j$, respectively.  For electrostatic interactions, the charge on the ionic species (PE monomers and counterions) is assumed to have a Gaussian charge density, with a spread $a_i$ chosen to reproduce the Born energy \cite{Wang2010FluctuationEnergy}; this results in a modified Coulomb interaction potential: $u_{elec}(r_{ij})=\frac{z_i z_j e^2}{4\pi \epsilon_0 \epsilon_r r_{ij}} \mbox{erf} \left ( \frac{\pi^{1/2}}{2^{1/2}a_{ij}r_{ij}}\right )$, where $z_i$ is the valence of bead $i$ and $\epsilon_0$ and $\epsilon_r$ are the vacuum permittivity and relative permittivity of the media, respectively. Similar to mass smearing, $a_{ij}=\sqrt{a_i^2+a_j^2}$. In this work, we choose $\sigma_i=\sigma_j=a_i=a_j=\sigma$ for all the particles, with $\sigma$ serving as the length unit in our simulation,  corresponding to about $0.6 \mbox{nm}$ from our coarse-graining \cite{Ye2024GCMe:Systems}. The energy unit is taken as the thermal energy $k_BT$ at room temperature, and the time unit is given by $\tau=\sqrt{m\sigma^2/k_BT}$ where $m$ is the mass of a bead assumed to be the same for all the particles. With $m$, $\sigma$, and $k_BT$ being the units of mass, length, and energy, respectively, the time unit $\tau$ is also 1. Unless otherwise stated, the polyelectrolytes are modeled as linear chains with unit-charged beads connected by a harmonic bond potential $E_{\mbox{bond}}=\frac{1}{2} K_{\mbox{bond}}(r-r_{0})^{2}$, with $K_{\mbox{bond}}=100$ and $r_0=0.5$. Equal amount of oppositely charged monovalent counterions are included in the simulation box to maintain charge neutrality. We set  $A_{ij}=25.2$ for all species to produce good solvent condition for the polymer. The relative permittivity is set as the water dielectric constant at room temperature. The simulation box has dimensions of $60 \times 60 \times 60$ containing a total of $540,000$ polyelectrolyte, counterion, and solvent beads. Other simulation details are given in Ref.\cite{Ye2024GCMe:Systems} and in the Supplemental Material. 

To incorporate hydrodynamic interactions, we use the pairwise Dissipative Particle Dynamics (DPD) thermostat, which is a Galilean-invariant thermostat that conserves local momentum and reproduces the correct hydrodynamics \cite{Espanol1995StatisticalDynamics,Ripoll2001LargeDynamics,Espanol2017Perspective:Dynamics}. The DPD thermostat contains an dissipative force and a random force given by\cite{Espanol1995StatisticalDynamics} $F_{ij}^D=-\alpha w_D(r_{ij}) (\bm{e_{ij} \cdot \bm{v_{ij}}})$ and $F_{ij}^R=\gamma w_R(r_{ij})\theta_{ij}\bm{e_{ij}}$, where $\bm{e_{ij}}=|\bm{r_i}-\bm{r_j}|/r_{ij}$ is the force direction and $\bm{v_{ij}=\bm{v_i}-\bm{v_j}}$ is the relative velocity between two beads. In the random force, $\gamma$ is related to the friction by $\gamma^2=2k_BT \alpha$, and $\theta$ is a random variable satisfying $\langle \theta_{ij}(t) \rangle =0$ and $\langle \theta_{ij}(t) \theta_{kl}(t')\rangle =(\delta_{ik}\delta_{jl} + \delta_{il}\delta_{jk}) \delta(t-t')$. The functional form of $w_D(r_{ij})$ and $w_R(r_{ij})$ can be chosen arbitrarily subjected to fluctuation-dissipation theorem that requires $w_R^2=w_D$ \cite{Espanol1995StatisticalDynamics,Groot1997DissipativeSimulation}. To be consistent with the spirit of our Gaussian-core model, here we choose $w_D=\exp(-\frac{3}{2\sigma_{ij}^2}r_{ij}^2)$. In our simulation, we set $\alpha=4.5$, which reproduces similar friction effect as in the classical DPD simulation \cite{Groot1997DissipativeSimulation}. We note that other thermostats commonly employed in molecular dynamics simulations, such as the Langevin thermostat and Nos\'e--Hoover thermostat, do not capture the correct hydrodynamics due to their failure to conserve momentum, even when explicit solvents are included \cite{Dunweg1993MolecularScreening,Stoyanov2005FromThermostat,Pastorino2007ComparisonSystems}.  

\begin{figure}
\centering
\includegraphics[width=.95\linewidth]{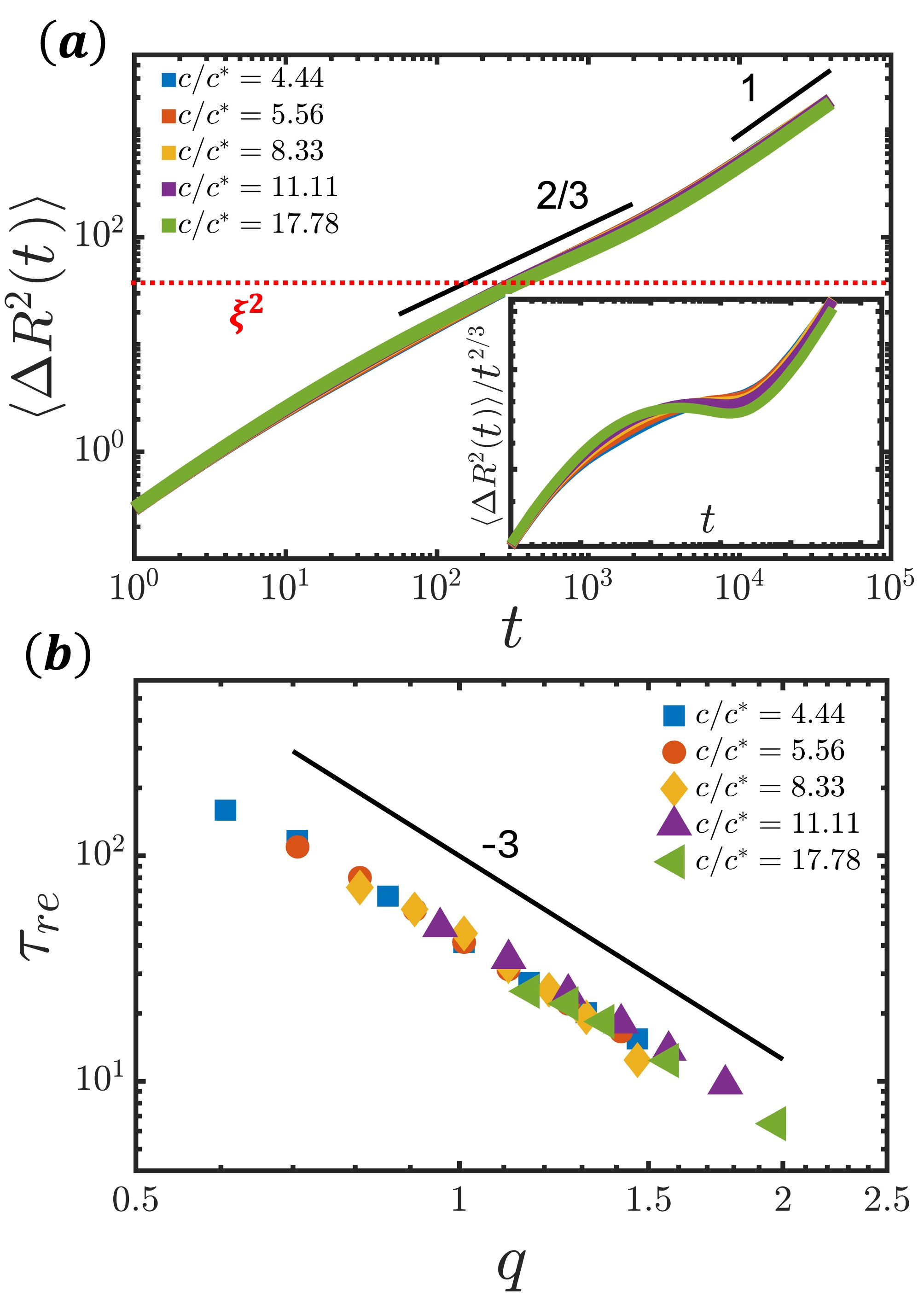}
\caption{(a) Monomer MSD in semidilute PE solutions with chain length $N=100$ at different polymer concentration $c/c^*=4.44,5.56,8.33, 11.11, 17.78$. The dashed line shows the square of the correlation length for the system $c/c^*=11.11$. The inset shows the MSD scaled by $\tau^{2/3}$. (b) Relaxation time of the polymer collective dynamics extracted from the intermediate scattering function.}
\label{fig:p1}
\end{figure}

We first examine the effects of hydrodynamics in semidilute PE solutions with chain length $N=100$ over a range of polymer concentrations $c/c^*=4.4 \sim 17.8$, where $c^*$ is the overlap concentration defined in Ref. \cite{Bollinger2021OverlapSolutions} that signals the onset of the semidilute regime. Figure 1(a) shows the monomer mean-squared distance (MSD) $\langle \Delta R^2 (t) \rangle$ of the center 4 beads of the PEs. Clearly, the subdiffusive regime exhibits Zimm-like dynamics: the MSD scales nearly as $~t^{2/3}$, as further evidenced by the appearance of an approximate plateau in the normalized MSD shown in the inset. Scaling theories \cite{Rubinstein1994DynamicsSolutions,Dobrynin1995ScalingSolutions} posit that hydrodynamics is screened beyond the correlation length $\xi$ as commonly defined by the position of structure factor peak of the PE solutions. In Fig. 1(a), we observe that the influence of Zimm dynamics goes well beyond the correlation length $\xi$ (see Supplemental Material for the calculation of $\xi$); subdiffusion is obviously more Zimm-like than Rouse-like before transiting to the diffusive regime. While scaling theories \cite{Rubinstein1994DynamicsSolutions,Dobrynin1995ScalingSolutions} focus on dynamics on length scales much larger than the correlation length, our results demonstrate the importance of hydrodynamics in the vicinity of the correlation length and beyond.  This information has not been provided in previous simulation studies because of the lack of hydrodynamic interactions \cite{Chang2002BrownianSolutions,Liao2007RouseStudy,Carrillo2011PolyelectrolytesSimulations}.   

The importance of hydrodynamics is also reflected in the collective dynamics of the polymer chains. For length scales below the correlation length, the intermediate scattering function can be fitted into a stretched exponential to obtain the relaxation time $\tau_{re}$ as a function of the wavenumber $q$ (see details in the Supplemental Material). In Fig. 1(b), we show the log-log plot of $\tau_{re}$ vs. $q$ for the different concentrations. The $\tau_{re} \sim q^{-3}$ scaling is characteristic of Zimm dynamics \cite{Doi1988TheDynamics}.  For $q > \xi^{-1}$, the collective dynamics of a section of the chain is the same as the single-chain dynamics \cite{Doi1988TheDynamics}. From scaling analysis, we have $\tau_{re} \sim q^{-2}/D$, where $D$ is the diffusivity of a section of the chain corresponding to the size $q^{-1}$. Invoking the Stokes--Einstein--Sutherland relation, $D\sim k_B T q /\eta$, where $\eta$ is the local viscosity, we then have $\tau_{re} \sim \eta q^{-3}/k_B T$.  That the data for the different concentrations all fall on the same curve indicates that the local viscosity $\eta$ is independent of concentration. This conclusion is also consistent with the long-time behavior shown in Fig. 1a, where the monomer diffusivity similarly shows independence of the PE concentration. The concentration independence of the chain diffusivity in the semidilute regime is in accord with theoretical predictions \cite{Rubinstein1994DynamicsSolutions,Dobrynin1995ScalingSolutions,Muthukumar1997DynamicsSolutions,Colby2010StructureSolutions} and experimental results \cite{Oostwal1993ChainPolystyrenesulfonate}.


\begin{figure}
\centering
\includegraphics[width=.95\linewidth]{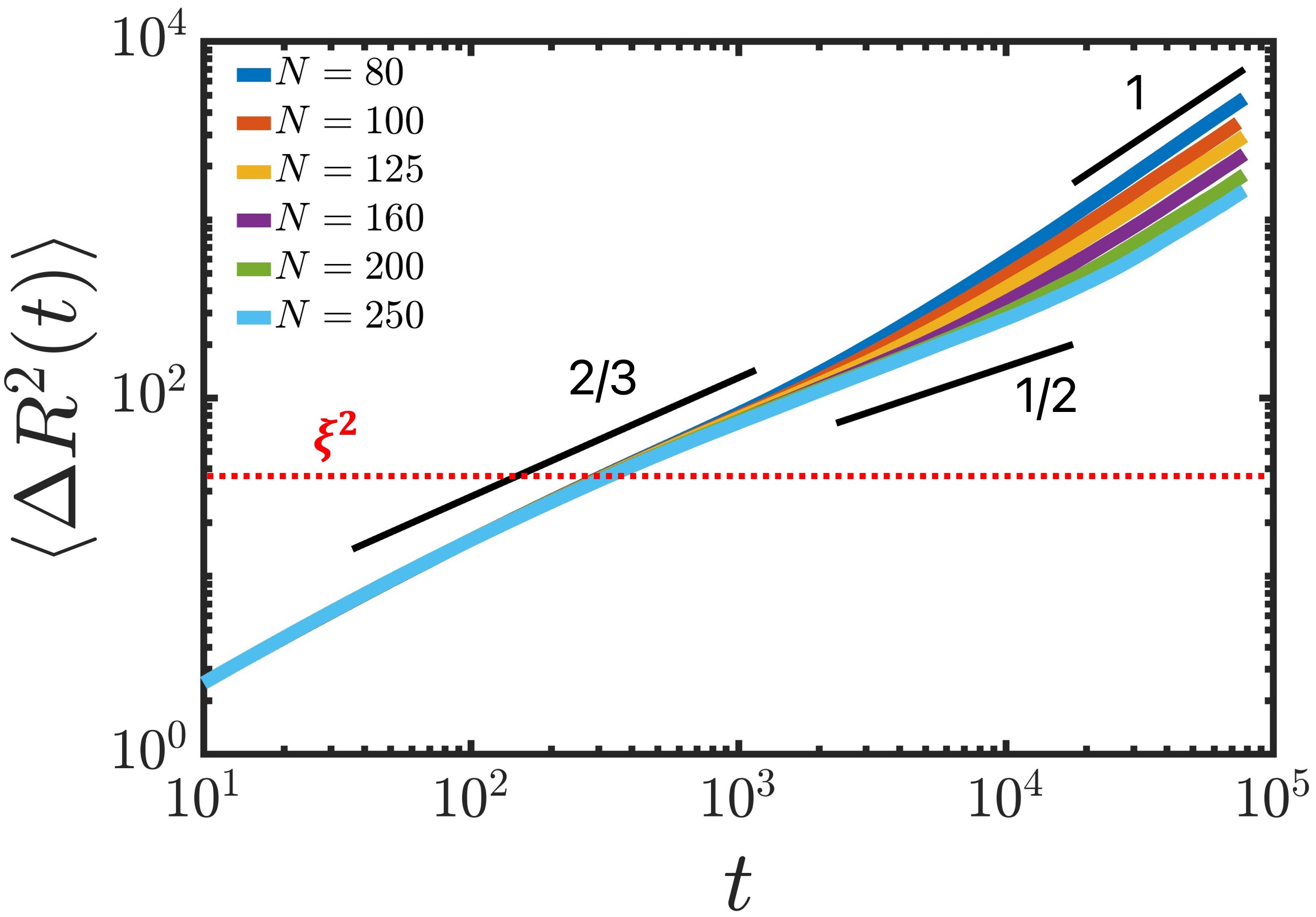}
\caption{Monomer MSD in semidilute PE solutions with fixed polymer density $c/c_{N=100}^*=11.1$ for different chain length $N=80 \sim 250$. The dashed line marks the value of the square of the correlation length for the $N=100$ system.}
\label{fig:p2}
\end{figure}

To examine the chain length dependence, in Figure 2(a) we show the monomer MSD in the semidilute regime for different chain lengths at a fixed monomer concentration of $c/c_{N=100}^* \approx 11.1$, corresponding to $3.7\%$ monomer volume fraction. We see that up to the length scale of the correlation length, the monomer MSD data for the different chain lengths completely overlap with each other and exhibit clear Zimm dynamics. For the shorter chains ($N \le 125$), the Zimm dynamics persists well beyond the correlation length and dominates the entire subdiffusive regime.  As chain length increases, there is deviation from the Zimm behavior towards the Rouse-like behavior in the subdiffusive regime, before transitioning to the diffusive regime at long times.  The results for the long-time center-of-mass chain diffusivity $D$ are provided in Fig. S2 in the Supplemental Material, where it is shown $D \sim N^{-1}$, in agreement with the theoretically predicted behavior \cite{Rubinstein1994DynamicsSolutions,Dobrynin1995ScalingSolutions,Muthukumar1997DynamicsSolutions}.  Together with the results shown in Fig. 1, we conclude that hydrodynamics dominates the subdiffusive regime in semidilute PE solutions of moderately short chains, even for concentrations one order of magnitude beyond the overlap concentration $c^*$. Since semidilute PE solutions of moderately short chains are relevant to many experimental studies and applications \cite{Izzo2014TheSolutions,Dolce2017IonizationMatters,Hu2020RecentApplications}, and most simulation studies employ moderately short chains, our results highlight the importance and necessity of properly including hydrodynamics in studying PE solutions in the semidilute regime. 

\begin{figure}
\centering
\includegraphics[width=.98\linewidth]{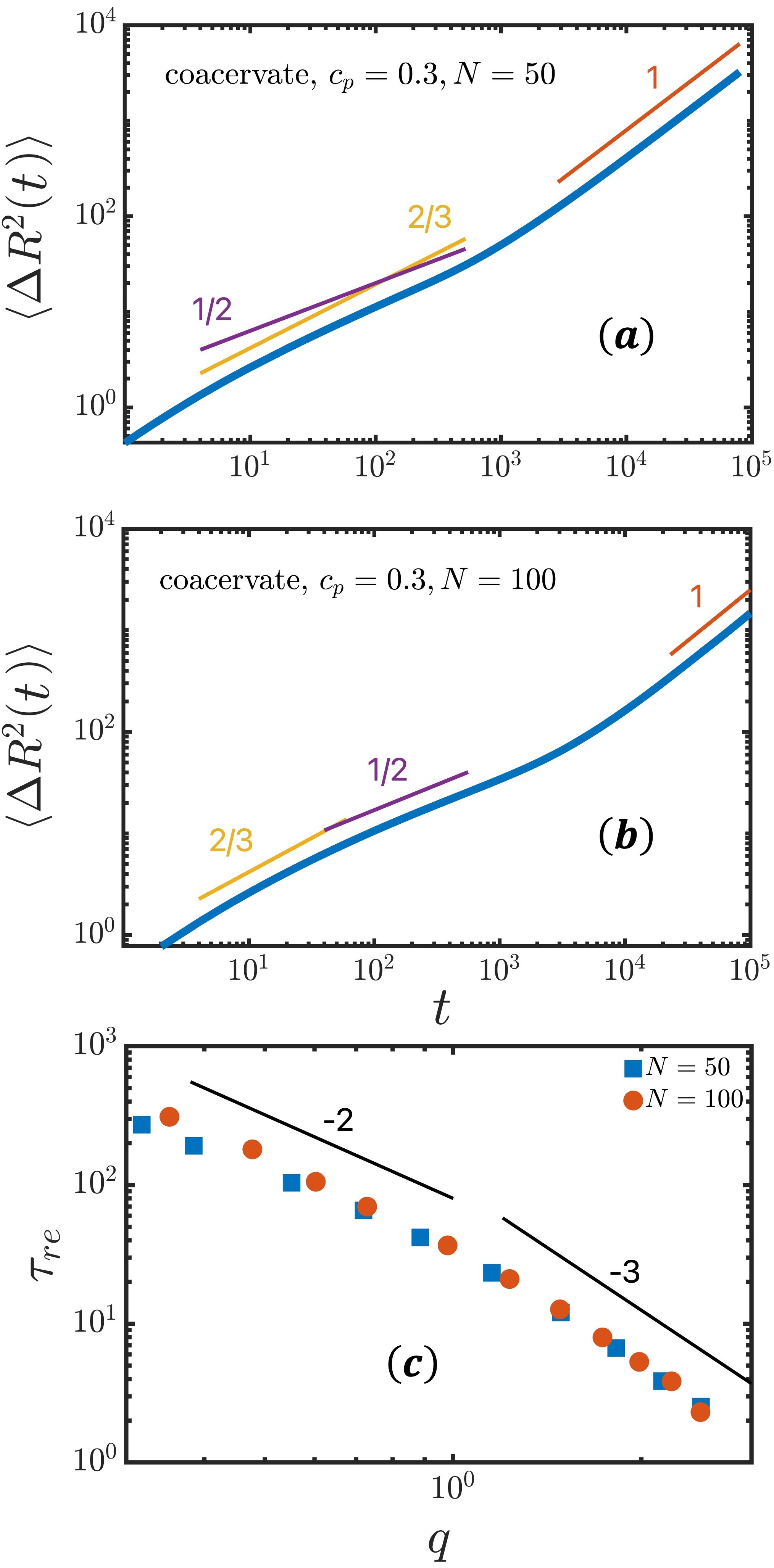}
\caption{Monomer MSD in a symmetric PE complex coacervate with $30 \%$ polymer concentration (a) PE chain length $N=50$. (b) PE chain length $N=100$. The $2/3$ and $1/2$ slopes represent Zimm and Rouse scaling, respectively. (c) Relaxation time of the polymer collective dynamics in the coacervate extracted from the intermediate scattering function.}
\label{fig:p3}
\end{figure}

Another class of systems that involve a high fraction of moderately short polyelectrolytes are the complex coacervates  formed as a result of liquid--liquid phase separation. For example, polypeptide-based coacervates usually contain $10\% \sim 30\%$ short polypeptides with chain lengths $N=5 \sim 100$ \cite{Koga2011PeptidenucleotideModel,Perry2015Chirality-selectedComplexes,Aumiller2016Phosphorylation-mediatedOrganelles,Abbas2021Peptide-basedProtocells,Sathyavageeswaran2024Self-AssemblingCoacervation}. 
Given the importance of the polymer dynamics in the coacervate properties such as aging \cite{Jawerth2020ProteinFluids}, viscoelasticity \cite{Syed2020TimeIonicComplexes}, and guest-molecule uptake and release \cite{McCall2018PartitioningCoacervates}, it is of interest to know the effects of hydrodynamics in these dense coacervate systems. 

To this end, we simulate two complex coacervate systems ($N=50$ and $N=100$) by mixing oppositely charged PEs with symmetric chain lengths.  For $N=50$, we obtain a coacervate (in coexistence with a supernatant solution) with about $21\%$ PEs, $77\%$ solvent, and $1-2 \%$ small ions. For $N=100$, the coexisting coacervate composition is $22\%$ PEs, $76\%$ solvent, and $1-2 \%$ small ions. To make comparison at the same composition, we choose a PE concentration above the coexistence limit: $30 \%$ PE, $68 \%$ solvent, and $2 \%$ small ions. The simulation details are given in the Supplemental Material. Figure 3(a) shows that for $N=50$, the subdiffusive behavior is completely dominated by Zimm dynamics, highlighting the importance of hydrodynamics. For the longer chain length $N=100$, the Zimm dynamics is followed by a transition to Rouse dynamics; see Fig. 3(b). Hydrodynamics also influences the collective dynamics of the coacervates, as shown in Fig. 3(c), in which we plot the relaxation time extracted from the intermediate scattering function for both PE species. For wavenumbers $q > 1$,  where 1 corresponds roughly to the length scale of about $6.3$ that is close to the radius of gyration of the chains ($8.9$ and $11.6$, respectively, for the two chain lengths), hydrodynamics completely dominates the collective dynamics.  The slope crosses over from $-3$ to $-2$ with decreasing wavenumber; the $-2$ slope at small $q$ is indication that the long length-scale dynamics reaches the diffusive regime.  
\begin{figure}
\centering\includegraphics[width=.95\linewidth]{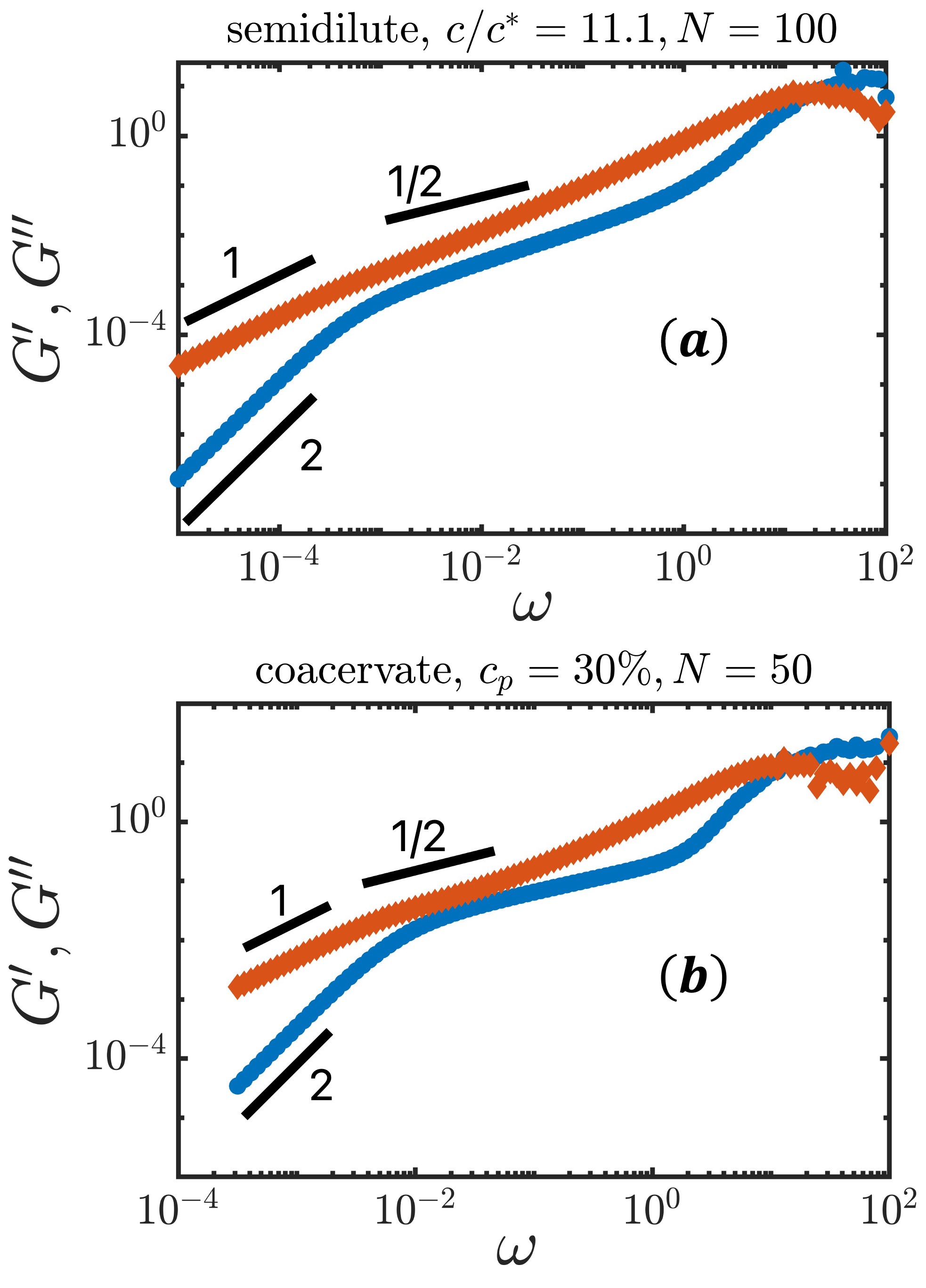}
\caption{Storage and loss moduli of (a) a semidilute PE solution, and (b) a PE complex coacervate. The $1/2$ scaling is the expected scaling from Rouse dynamics (for $G'$).}
\label{fig:p4}
\end{figure}

The effects of hydrodynamics on chain dynamics at intermediate time and length scales will impact the rheological response. We expect the hydrodynamic effects to be manifested in two ways: first, they should directly affect the stress relaxation behavior on this time scale, and second, they determine the effective monomer friction of the chain on longer time scales even when the chain dynamics becomes Rouse-like. We note that Zimm dynamics and Rouse dynamics result in nearly indistinguishable scaling behavior in the stress relaxation \cite{Rubinstein2003PolymerPhysics}. Regardless, a fully consistent calculation of the rheological response needs to account for the hydrodynamic interaction. In Figs. 4(a) and 4(b) we show the dynamic moduli of a semidilute PE solution and of a coacervate, respectively. Here, the storage module  $G'$ and loss module $G''$ are derived, respectively, from the sine and cosine transform of the stress relaxation function of the system \cite{Daivis1994ComparisonDecane,Ramirez2010EfficientSimulations,Thompson2022LAMMPSScales} (for details, see the Supplemental Material). In the low frequency ($\omega$) regime, $G'$ and $G''$ scale respectively as $\sim \omega^2$ and $\sim \omega^1$ in both systems, characteristic of viscoelastic polymeric liquids. Deviation from the $\sim \omega^2$ behavior for $G'$ sets in around $\omega^{*} \sim 10^{-3}$ and $\omega^{*} \sim 10^{-2}$, respectively, for the two systems. The corresponding time scales ($\sim 2 \pi /\omega^{*}$) are consistent with the crossover between the subdiffusive and diffusive dynamics for the monomer diffusion, shown in Fig. 2 and Fig. 3.  Because the system contains mostly solvent, the solvent contribution to the the loss modulus makes it difficult to see a clear change in the behavior of $G''$ at $\omega^{*}$ \footnote{Some authors present the loss modulus as $G''-\omega \eta_s$ by subtracting the solvent contribution \cite{Rubinstein2003PolymerPhysics}. However due to the large error bars in the difference between the solution stress relaxation function and the solvent relaxation function, we are unable to obtain reliable data for this quantity.}.

In summary, using molecular dynamics simulation that explicitly accounts for hydrodynamics, we have systematically studied the polymer dynamics in semidilute PE solutions and PE complex coacervates. For moderately short chains, our results show that hydrodynamics dominates the chain dynamics in semidilute PE solutions well beyond the correlation length. Hydrodynamics also dominates the chain dynamics at intermediate time and length scales in complex coacervates consisting of moderately short chains, which are relevant to many polypeptide-based biocondensates. For both semidilute PE solution and PE coacervate systems with longer chains, a Zimm to Rouse crossover is observed. Our study fills a gap in the existing simulation literature on the dynamics of semidilute PE solution and PE coacervate systems by highlighting the importance of including hydrodynamics in the study of these dense PE solutions.  We expect that hydrodynamics will have important consequences for many properties in PE solutions, such as charge transport \cite{Grass2008ImportanceChains,Frank2009MesoscaleFields} in both the linear and nonlinear response regimes, and nonlinear viscoelastic behavior.  


\hfill
\begin{acknowledgments}
This research is supported by funding from Hong Kong Quantum AI Lab,
AIR@InnoHK of Hong Kong Government. We thank the general computation time allocated by the resources of the Center for Functional Nanomaterials (CFN), which is a U.S. Department of Energy Office of Science User Facility, at Brookhaven National Laboratory under Contract No. DE-SC0012704. 
\end{acknowledgments}

\bibliography{hydro}

\end{document}